\documentclass[amsmat,amssymb,amsfonts,aps,prb,twocolumn,showpacs]{revtex4}

\usepackage{graphicx}
\usepackage{dcolumn}
\usepackage{bm}
\begin{document}

\title{Spin-filtered edge states in graphene}

\author{ D. Gos\'albez-Mart\'inez(1), D. Soriano(1,2), J. J. Palacios(3), J. Fern\'andez-Rossier (1,4)}
\affiliation{(1) Departamento de F{\' i}sica Aplicada, Universidad de Alicante, San Vicente del Raspeig, 03690 Alicante, Spain. 
\\ (2) CIN2 (ICN-CSIC) and Universitat Aut\`onoma de Barcelona, Catalan Institute of Nanotechnology, Campus UAB, 08193 Bellaterra (Barcelona), Spain
\\
   (3) Departamento de F\'isica de la Materia Condensada, Universidad Aut\'onoma de Madrid, Cantoblanco,28049 Madrid, Spain.
\\ (4) Iberian International Nanotechnology Laboratory, Avenida  Mestre Jos\'e Veiga, 4715-330  Braga, Portugal}

\date{\today} 

\begin{abstract} 

Spin orbit coupling changes  graphene, in principle, into a two-dimensional topological insulator, also known as quantum spin Hall insulator. 
One of the expected consequences is the existence of spin-filtered edge states that 
carry dissipationless spin currents and  undergo  no back-scattering  in the presence of non-magnetic disorder, leading to quantization of conductance. Whereas,
due to  the small size of spin orbit coupling in graphene, the  
experimental observation of these remarkable predictions is unlikely,  the theoretical understanding of 
these spin-filtered states is shedding light on the electronic properties of edge states in other two-dimensional quantum spin Hall insulators.  
Here we review the effect of a variety of perturbations, like curvature, disorder, edge reconstruction,  edge crystallographic orientation,
 and Coulomb interactions  on the electronic properties of these spin filtered states.  

\end{abstract}

\maketitle
\section{Introduction}

Since the fabrication of high mobility graphene based transistors, back in 2004\cite{Novoselov04}, the exploration of graphene has led to a number of striking  experimental findings\cite{RMP09}  such
as the quantum Hall effect \cite{Novoselov05,Zhang05}, even at room temperature\cite{ }, the Klein tunneling, the universal optical absorption, and also to an even larger amount of  theory work  predicting   
exotic electronic phases in graphene.  Whereas most of the new electronic phases in Condensed Matter Physics are driven by Coulomb interactions,  in 2005 Kane and Mele (KM)
predicted that spin orbit coupling (SOC) alone would turn graphene into a completely new electronic state, the so-called quantum spin Hall insulator (QSHI)\cite{Kane-Mele1,Kane-Mele2}.
  Inspection of the  energy bands of graphene, as described within the Kane-Mele model, look deceptively like an ordinary narrow gap semiconductor. However, a careful study of
 the wave function topology in reciprocal space reveals that the gap opened at the Dirac points by the SOC has different signs at the two valleys. 
The twisting  in the reciprocal state makes graphene topologically different from vacuum or graphene without SOC. The most conspicuous consequence of the topologically 
non-trivial band structure of the QSHI is the existence of spin-filtered edge states at the boundaries of graphene with other topologically trivial systems.

The quantum spin Hall phase proposed by Kane and Mele can be portrayed as two copies of a quantum Hall phase\cite{Kane-Mele2}, one for each spin orientation and with opposite spin-dependent  
field orientations. Like in the conventional quantum Hall phase, the bulk states are gapped, but the edges hold states at the  Fermi energy that  carry current. In the quantum spin
Hall phase, at a given edge, spin up electrons flow in the opposite direction to that of spin down electrons, resulting in a pure spin current flow.   
Since  backscattering in a given edge requires reversal of the spin,  the conductance of the edge states is expected to be quantized even in the presence of 
time-reversal preserving disorder.  

In their original work,   KM studied the competition between the intrinsic SOC, which drives the QSHI phase, and inversion symmetry breaking perturbations, such as
 a Rashba spin orbit term and a staggered perturbation, which make graphene an ordinary insulator.   The interplay of these couplings with non-magnetic disorder 
was later addressed by L.Sheng {\em et al.}\cite{Haldane05} and D.N.Sheng {\em et al.}\cite{Haldane06b}. They found that, as expected, the  QSHI phase, and thereby the spin-filtered edge states, 
were robust with respect to the addition of time-reversal symmetric disorder.  
   
Here we address four different topics: first, the relation  of the KM model with the microscopic description of SOC in graphene. 
In section II, we show that the KM Hamiltonian, originally derived from heuristic symmetry considerations,   is a good effective one orbital description of 
the low energy bands of flat graphene, as described with a 4-orbital Slater-Koster (SK) 
Hamiltonian including SOC\cite{Min06,Huertas06,Yao07,Fabian2010}.  

Our second topic refers to the influence of the geometry of the sample on the the electronic properties  of the spin-filtered edge states. We consider 
the effect of the crystallographic orientation, either zigzag or armchair, the effect of a reconstruction into the so-called reczag edge, and the effect of the curvature. 
   In the original papers, KM studied the spin-filtered states in graphene terminated by zigzag boundaries. 
 As we discuss in sections \ref{terminations} and \ref{Hubbard}, the choice of zigzag ribbons with pre-existing non-spin filtered edge states, 
obscures somewhat the dramatic effect of SOC on the electronic properties of graphene.

 The third topic addressed in section \ref{disorder} is related to the robustness of the edge state conductance with respect to the presence  of non-magnetic disorder. 
We study the effect of various types of disorder, Anderson type and edge roughness, on the two terminal conductance of zigzag terminated ribbons.
 
Finally,  our fourth topic concerns the study of the robustness of the edge states with respect to magnetic order. In the case of zigzag ribbons ignoring SOC, 
the edge states undergo a Stoner inestability\cite{Fujita96,Son06,Cohen06,JFR08} for arbitrary small electronic repulsion. Thus, the robustness of the 
spin-filtered edge states is compromised\cite{Soriano-JFR10}, unlike in the case of armchair terminations.    
In section IV we discuss the interplay between SOC and electron-electron interactions in both geometries.

\section{Spin orbit in graphene and the Kane Mele Hamiltonian}
\label{section2}

The KM Hamiltonian can be seen from two different perspectives:  as a theoretical model that describes a new phase of matter, the QSHI  phase, or as a 
effective model Hamiltonian to describe the effect  of SOC on the   $\pi$ electrons of graphene. The KM model  is inspired by a mathematical model for spinless 
fermions under the influence of an inhomogeneous magnetic field that, with a vanishing total flux, gave rise to a quantized Hall response. 
This  model, originally proposed by Haldane\cite{Haldane88}, breaks time reversal invariance but shows a way to obtain quantized Hall effect without an applied external field.  
By including the spin degree of freedom and  giving to the inhomogeneous magnetic field  a spin dependent value, KM found a way to restore time-reversal symmetry in the model, 
which would yield the so called QSHI phase. 

\subsection{Spin orbit coupling in graphene}

We briefly discuss the fundamentals of SOC in graphene. 
The low energy physics of graphene, ignoring SOC,  is described by the $\pi$ orbitals, with orbital angular momentum $l=1$ and $m=0$,
 if we chose the quantization axis perpendicular to the plane. 
The SOC, as described by

\begin{equation}
\hat{H}_{SO}=\lambda \sum_{i}\vec{L}_i \cdot \vec{s}_i 
\label{SOC}
\end{equation}

couples, to lowest order in $\lambda$,  the $\pi$ electrons with spin $s_z$ to $l=1,m=\pm 1$ electrons with opposite spin $\overline{s}_z$.   Thus,  SOC 
mixes the $\pi$ electron with the $\sigma$ electrons.  A possible approach is to give up an effective description in terms of the $\pi$ orbitals only and use a 
tight-binding model that includes the four atomic orbitals, i.e., the three responsible for the $sp^2$ hybridization and the $\pi$ orbital, in each atom. Thus,  
there are 8 spin orbitals per carbon atom and a total of 16 in the unit cell of the honeycomb lattice.  In this case,  the  16 x 16 Hamiltonian can be obtained 
using a SK parametrization of the tight-binding Hamiltonian\cite{Min06,Huertas06,Dani2011}.  In figure 1 (left panel) we show    the energy bands of graphene so obtained. 
For the sake of clarity we use an  unrealistically large SOC, $\lambda=4$eV, 3 orders of magnitude above the accepted value $\lambda\simeq 10 meV$.   
Whereas the large energy scale is not much affected by the SOC, a band gap opens right at the $K$ and $K'$ points (right panel) . 
For realistic values of the SOC, $\lambda\simeq 10 meV$, the value of the gap is in the range of 10$\mu eV$. Thus, 
from a practical perspective, the  SOC gap in graphene can be ignored in most instances.  

\begin{figure}
[t]
\includegraphics[width=0.90\linewidth,angle=0]{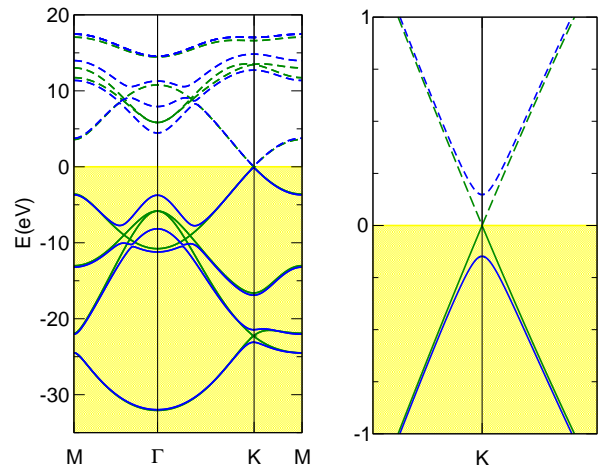}
\caption{ \label{figure1}
(\emph{Color online}) Left panel: Graphene band structure computed with the 4-orbital orthogonal tight-binding model with a SOC of  $\lambda$=4eV (blue lines) and without SOC 
(green lines). Dashed and solid lines represent empty and occupied bands, respectively. Right panel: zoom of the band structure at $K$ point.}
\end{figure}

Within this 4-orbital picture, the smallness of the band gap  comes from the fact that $\lambda$ is small and it only affects the $\pi$ bands to second order\cite{Min06,Huertas06}. 
Although it does not make a difference in the rest of the discussion, it is worth pointing out that the 4-orbital picture is not complete because the $d$ orbitals happen to 
play a role.  The $xz$ and $yz$ orbitals are weakly hybridized to the $\pi$ orbitals, even for zero SOC.  The $d$ orbitals, in turn, have their own SOC, 
as a result of which the  $\pi$-$d$ mixing is spin dependent. This results in a contribution to the $\pi$ bands with the same symmetry than the one coming from 
the $\lambda^2$  coupling through the $\sigma$ bands, but twenty times higher \cite{Fabian2010}.

\subsection{The Kane Mele Hamiltonian}
\begin{figure}
[t]
\includegraphics[width=0.80\linewidth,angle=0]{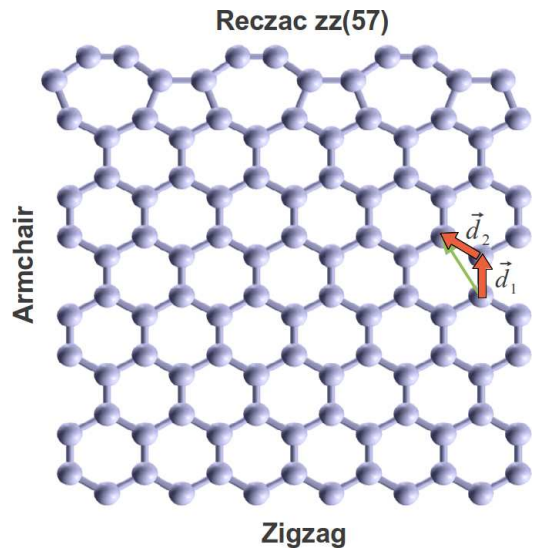}
\caption{ \label{KM-term}
(\emph{Color online}).  The three types of edges considered: reczag (top), armchair (sides), and zigzag (bottom). 
The bond-vectors $\vec{d}_i$ featured in the Kane-Mele Hamiltonian are also shown. }
\end{figure}

The results above suggest that some sort of effective Hamiltonian where  the higher energy bands are integrated out can describe the low energy physics 
around the $K$ points in terms of the $\pi$ orbitals only.   The numerical calculations and symmetry considerations show that close to the $K$ points 
$s_z$ is a good quantum number.  In the lattice representation, the effective model that does the job was proposed by KM:
\begin{equation}
\label{KMHamiltonian}
H_{KM}=t_{\mathrm{KM}}\sum_{\langle\langle i,j \rangle\rangle} c_i^\dagger \mathbf{\sigma} (\mathbf{d}_{kj} \times \mathbf{d}_{ik}) c_j  
\end{equation}
where the  SOC yields   a spin-dependent term which connects second neighbors in the honeycomb lattice through the two vectors $\mathbf{d}_{kj}$ and $\mathbf{d}_{ik}$ as shown in o
figure (\ref{KM-term}), and $t_\mathrm{KM}$ is the spin-orbit coupling strength. 
When summed to the standard first neighbor hopping $t$ term, the 
 momentum representation of  the Hamiltonian for graphene, including SOC, can be written as: 
\begin{equation}
{\cal H}(\vec{k})= t_{\rm KM}  g(\vec{k}) s_z \sigma_z +  t f(\vec{k}) \sigma_x 
\label{KM-2Dgraphene}
\end{equation}
where $\sigma_x$ and $\sigma_z$ are the Pauli matrices in the sub lattice space, $t$ is the first neighbor hopping, $f(\vec{k})=1+Exp[i\vec{k}\cdot\vec{a}_1]+Exp[i\vec{k}\cdot\vec{a}_2]$, $g(\vec{k})= i(-Exp[i\vec{k}\cdot\vec{a}_1]+Exp[i\vec{k}\cdot\vec{a}_2]-Exp[i\vec{k}\cdot\vec{a}_3]+Exp[i\vec{k}\cdot\vec{a}_1]-Exp[i\vec{k}\cdot\vec{a}_2]+Exp[i\vec{k}\cdot\vec{a}_3])$ and $t_{KM}$ is the spin-orbit coupling strength.  

Notice that $g(\vec{k})=-g(-\vec{k})$ because  the sign of the spin-dependent second-neighbor hopping is also sensitive to the clockwise-anticlockwise nature of the hopping path.
At the Dirac points $f(\vec{k})$ vanishes and the effect of the SOC Hamiltonian is to open a valley and spin dependent gap given by the following Hamiltonian:
\begin{equation}
{\cal H}_{SO}= \Delta_{SO} \tau_z\sigma_z s_z
\label{KpointH}
\end{equation}
where  $\Delta_{SO}= 3\sqrt{3} t_{\rm KM}$  is the band gap divided by 2,
$\tau_z=\pm1 $ for $K$ and $K'$ valleys, $\sigma_z=\pm 1$ for $A$ and $B$ subltatice and $s_z=\pm 1/2$ for the spin projection perpendicular to the plane of graphene.  At the $K$ and $K'$ valley the top of the valence band and bottom of the conduction band are sub lattice polarized.   It is quite apparent that, for a given spin orientation $s_z=+\frac{1}{2}$, the wave functions change sub lattice character as we change the valley index. Thus, the gap changes sign, at a given spin, when we go from $K$ to $K'$.
This twist in the reciprocal space makes the SOC gap topologically  different from the gap opened
by a sublattice symmetry breaking or  staggered potential ${\cal H}_{stg}=\Delta\tau_z$ which favors the same sublattice at both valleys.

\section{Edge states: the effect of  crystallographic orientation }
\label{terminations}

Whereas the QSHI phase is a bulk property,  it is revealed at the boundaries through the appearance of edge states, 
very much like it happens in quantum Hall bars\cite{Halperin82}.
In most instances edges states are studied by  looking at sufficiently wide strips with two edges 
rather than  semi-infinite systems with only one edge.   Strips or ribbons are one-dimensional crystals whose electronic structure is 
simplified taking advantage of the translational invariance along the longitudinal (edge) direction, which we take to be along the $x$ axis.  
The existence of edge states is an expected consequence of the non trivial topological order of the QSHI phase. However, the study of their 
properties in ribbons of the honeycomb lattice has some subtleties associated both with inter-edge coupling, on account of their finite 
transverse  size, and with the existence, along some crystallographic directions,  of edge states even in the absence of SOC\cite{Nakada96}.  
Here we discuss  the edge states in three different types of termination:  zigzag, armchair and reczag, using the two different tight-binding
 models described above, namely, the effective 1-orbital KM and the 4-orbital SK model. We also study the effect of 
curvature on the electronic properties of the zigzag edges. 

\subsection{Zigzag edge states}
\begin{figure}
[t]
\includegraphics[width=0.90\linewidth,angle=0]{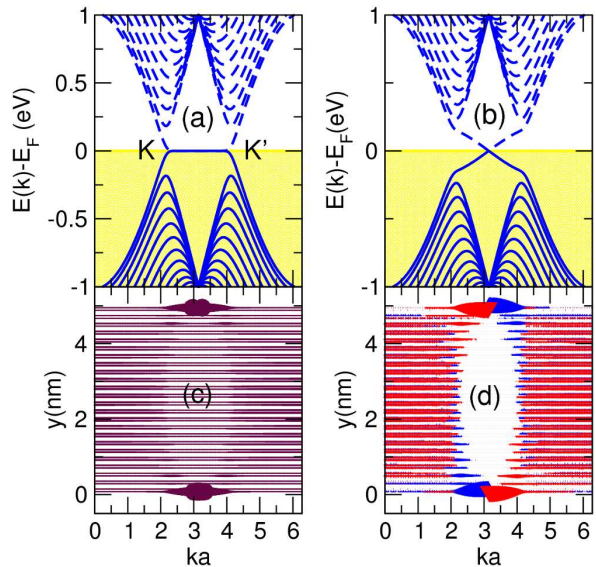}
\caption{ \label{figureIII}
(\emph{Color online}). Band structure of a graphene zigzag ribbon for (a) $t_{KM}/t = 0$ and (b) $t_{KM}/t = 0.03$. The unit cell 
contains $N=48$ atoms. Panels (c) and (d) represent the highest valence band charge density $|\Psi^{VB}_i|^2$ for each site $i$ across the 
ribbon width. Red and blue colors refer to different spin projections. When the SOC is turned on,
 the valence and conduction bands acquire some dispersion between $K$ and $K'$ points and become spin-filtered.}
\end{figure}

The most studied graphene termination is the so-called zigzag edge  
[shown in Fig. (\ref{KM-term})], which presents pre-existing edge states, i.e., edge states even without SOC.    
This can be seen in Fig. \ref{figureIII}, where  we plot the band structure of a zigzag ribbon with $N=48$ sites in each unit cell
calculated within the KM model (identical results are obtained within the SK model\cite{Dani2011}). 
We consider two cases, one without SOC and one with a finite SOC . 
In both cases the bands have a two-fold degeneracy associated with the spin. (In the case without SOC the degeneracy associated with
the existence of two identical edges is slightly removed by the interaction between edges.)  
The wave functions (modulus square) across the $y$ direction
of the highest energy valence band  as a function of the longitudinal momentum $k$ are shown in the lower panels. 
Without SOC, the flat mid gap bands correspond to localized states at the edges\cite{Nakada96}. In the case of zigzag ribbons,  
the valley character is preserved\cite{Brey-Fertig}.  Thus, it can be said that edge states go from $K$ to $K'$ in momentum space.   
In the presence of SOC (right panels)  the valence and conduction bands connecting $K$ and $K'$ points in the Brillouin zone acquire
 an apparently innocent dispersion, but the wave functions undergoes a dramatic change. Whereas
in both cases the states at the valence and conduction bands along the path $K \rightarrow K'$ are localized at the zigzag edges,
 in the presence of SOC they become spin-filtered, giving rise to helical edge states\cite{Kane-Mele1,Kane-Mele2}. 
This means that electrons with opposite(same) spin carry current in opposite direction along the same(opposite) edge.     
Interestingly, the occupied bulk states also carry a spin current in the opposite direction so that the net spin current at the edge 
is zero\cite{Wu2011}

\subsection{Armchair edge states}

\begin{figure}
[t]
\includegraphics[width=0.90\linewidth,angle=0]{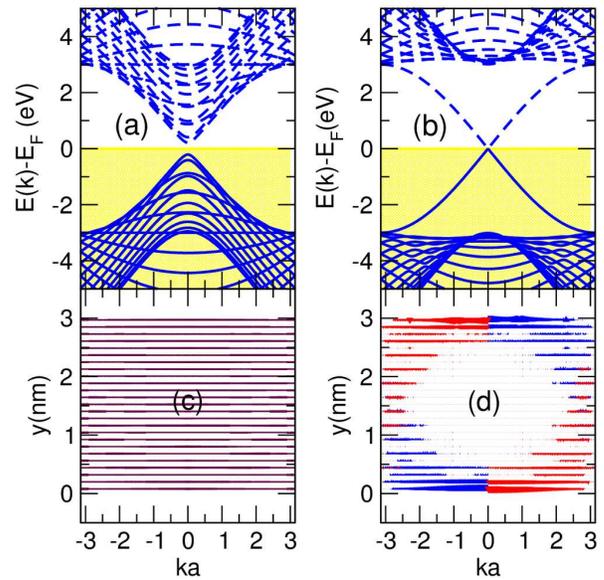}
\caption{ \label{figure5}
(\emph{Color online}) Band structure and squared highest valence band wavefunction $|\Psi^{VB}_i|^2$ for an armchair 
graphene ribbon with $N=50$ atoms across the ribbon width. Red and blue colors refer to different spin projections.
For panels (a) and (c), $t_{KM}/t=0$ and for (b) and (d), $t_{KM}/t = 0.1$.}
\end{figure}

We now consider a different crystallographic orientation of a honeycomb lattice, the armchair edges  [see Fig. \ref{KM-term}]. 
  In Fig. \ref{figure5} we plot the energy bands of the KM model for  a semiconducting armchair ribbon with $N=50$ sites in the unit cell, 
both with null (left) and finite (right) SOC.  In contrast to the zigzag ribbons, this termination lacks pre-existing edge states.  
Within this model and with zero SOC,  the band structure of armchair ribbons is either conducting or insulating depending on the 
width\cite{Nakada96,Brey-Fertig}.  In the lower panels we show the wave functions (modulus square) for each spin projection
of the highest valence band with no SOC.  It is apparent that they are delocalized along the transverse direction. 

By making the ribbon sufficiently wide,  the confinement  gap is smaller than the  SOC gap. When this happens, 
two bands crossing the Fermi energy appear whose wave functions are localized at the edges. In addition, their wave functions 
[Fig. \ref{figure5}(d)]  show the same correlation between velocity, spin orientation and edge than their counterparts in the zigzag ribbons.  
Thus, spin-filtered edge states appear without having preformed edge states. 
Similar results have been obtained by E. Prada {\em et al.} \cite{Prada2011}
 using the $kp$ approach, originally devised for ribbons without SOC\cite{Brey-Fertig}, complemented with the SOC term in Eq. (\ref{KpointH}).  

\subsection{Reczag edge states}

\label{termination}

We now consider reconstructed edge states in zigzag graphene ribbons. This is motivated by density functional theory calculations 
claiming that zigzag edges are prone to  edge reconstructions\cite{koskinen}. Here we consider the so-called reczag, 
which is a self-passivated zigzag edge, where the edge hexagons change into a configuration of alternating pentagon and heptagons. 
In order to study the electronic structure and the edge states of these systems, we make use of our 4-orbital orthogonal SK model 
with SOC, taking into account that the hopping depends on the distance\cite{Harrison}. This is the a necessary methodology here because a 
the 1-orbital KM model cannot account for the physics of the newly formed atomic structure. 
Density functional theory calculations including SOC are also valid, but may require in some cases great computational resources.

\begin{figure}
[hbt]
\includegraphics[width=0.90\linewidth,angle=0]{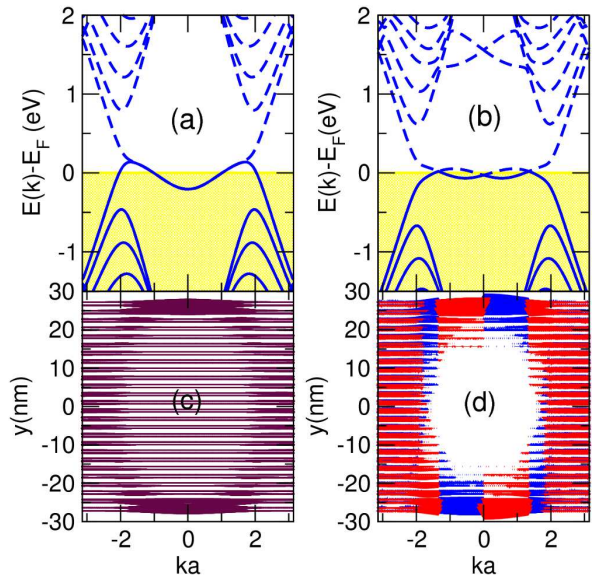}
\caption{ \label{RECZAG}
(\emph{Color online}) Band structure and squared highest valence band edge states wavefunctions $|\Psi^{VB}_i|^2$ for 
a reczac graphene ribbon with 104 carbon atoms per unit cell. For panels (a) and (c), $\lambda=$0eV and for (b) and (d), $\lambda =$ 4eV.}
\end{figure}

We obtain the geometrical disposition of the atoms from Ref. \onlinecite{Peres-reczac} and the SK parameters from Ref. \onlinecite{Dani2011}. 
The electronic configuration for the low energy spectra is shown in the Fig. \ref{RECZAG} both with null (left) and finite (right) SOC. 
Because of the reconstruction, the unit cell of the reczag ribbons is twice as large along the longitudinal direction and, thereby,
 the Brillouin zone is half as large as the one for the zigzag ribbons.  As a result, the edge states occupy a larger percentage of 
the Brillouin zone. 
It is apparent that  the band edges are no longer electron-hole symmetric. This is due to the loss of the bipartite character of 
the reconstructed lattice.  This also results in dispersive edge states, in contrast to the zigzag case.
 In the lower panels of Fig. \ref{RECZAG}  we show the wave function map for the lowest energy conduction band states 
projected on each atom as function of $k$. Without SOC,  the reczag ribbon presents edge states that are not spin-filtered and present
a four-fold degeneracy (edge and spin).  When the SOC is turned on, the four-fold degeneracy is lifted and they acquire a spin-filtered character, 
shown in Fig.  \ref{RECZAG}(d).  In this instance, the non-monotonic character of the band velocity makes the effect more intricate. 
As a  consequence of the shape of the bands, each of the two spin-filtered edge states cuts three times the  Fermi level, introducing 
more conduction channels near the Fermi energy.

\subsection{The effect of curvature}

The effect of SOC in graphene is small due, in part, to the decoupling of the $\pi$ and $\sigma$ bands for flat graphene. 
Since it is possible to have curved graphene ribbons, made, e.g., by opening carbon nanotubes \cite{DAI,unzipping}, the interest 
on the effect of curvature on the spin filtered edge states has increased\cite{Dani2011}.
In a curved ribbon,  the  $\pi$ orbitals are not longer decoupled from the $\sigma$ orbitals.  This is similar  to the case of flat 
graphene under the action of a perpendicular electric field\cite{Min06}.  As a consequence,  we can expect a
new terms in the effective hamiltonian for the $\pi$ linear in the atomic SOC constant $\lambda$, as a result of which $s_z$  
is not a good quantum number anymore.  In the case of flat graphene under an external electric field the new term is the so-called 
extrinsic SOC or Rashba interaction,  which acts as a spin-flipping nearest-neighbor hopping. 

\begin{figure}
[t]
\includegraphics[width=0.90\linewidth,angle=0]{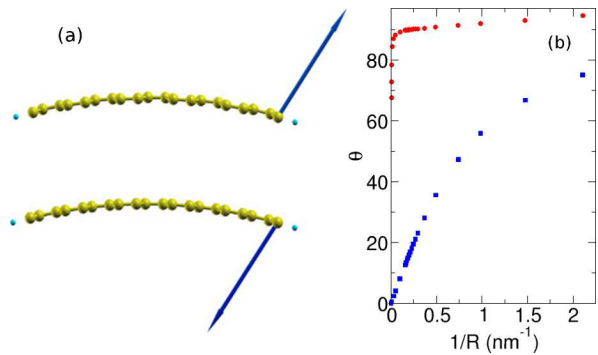}
\caption{ \label{figure4}
(\emph{Color online}) (a) Spin polarization of edge states at $k=0.99 \pi/a$(top) and $k=-0.99 \pi/a$ (bottom). (b) 
Angle between the normal direction of the edge and the spin polarization as a function of the curvature 
for $\lambda=5meV$ (red dots) and $\lambda=2eV$ (blue squares).}
\end{figure}

The electronic structure of curved graphene ribbons with zigzag terminations is shown in 
in Fig. \ref{figure4} and   has been obtain within the 4-orbital SK model including SOC\cite{Dani2011}.
The low energy band structure (not shown) remains similar to the case of flat ribbons, except for a the appearance of a small dispersion 
in the originally flat bands. For finite SOC, the edge bands are spin filtered as well, but with an important difference with respect 
to the flat case. Since  $s_z$ is not a good quantum number, the spin polarization of these states acquire an angle, 
$\theta$, with respect to the normal direction at each atom of the edge. In Fig. \ref{figure4}(a) we represent the wavefunction of the 
edge states at $k=\pm 0.99 \pi/a$.  The fact that the spin polarization points in opposite direction 
for these two  $k$ states  reveals they form a time reversal Kramers pairs. 
 In Fig. \ref{figure4}(b) we also show how  $\theta$ evolves with respect to  the curvature for two different values of the SOC.
   We observe that the effect of curvature is dramatic for a realistic value of 
the spin-orbit coupling, where almost negligible curvatures make the spin point in the tangential direction. 
   
   As a conclusion for this section, we have seen how SOC gives rise to the appearance of spin filtered edge states, 
regardless of the crystallographic direction of the edge.  Interestingly, the bands of the edge states do depend on the direction.  
Unfortunately, the energy scale for the edge states dispersion is below 1$meV$ when computed with realistic values of the SOC for graphene. 
 However,  understanding  spin filtered edge states in this chemically simple situation could shed some light on the edge states in 
more complicated systems.  

\section{Disorder and electronic transport}
\label{disorder}

The topological character of the QSHI phase manifests itself in the robustness of the spin-filtered edge states. 
In the presence of non-magnetic disorder they behave like a pair of robust quantum channels, presenting a conductance quantization, $G_0=2e^2/h$,
even when $s_z$ is not a good quantum number\cite{Kane-Mele1,Kane-Mele2}. Shortly after KM proposed the existence of a QSHI phase in graphene, 
the conductance quantization was numerically verified by L. Sheng {\em et al.} for the Anderson disorder model\cite{Haldane05,Haldane06b}. 
Here we carry out a similar study through an analysis of the transport properties of zigzag ribbons in a variety of situations. 
We are mostly intrigued by the case of curved ribbons as the ones introduced in previous section. First, for illustration purposes, 
we revisit the problem in flat ribbons. As shown in Sec.\ref{section2}, zigzag graphene ribbons with SOC constitute the most simple model of 
QSHI since they present a single branch of helical states for any width of the ribbon and only inter-edge backscattering could affect 
the quantization of conductance. Finally, we ask ourselves if the spin non-conserving effect of the curvature can play any role in 
the inter-edge backscattering.

We study two types of disorder:  Anderson type, where  we sum  random on-site values $\epsilon_i$ distributed uniformly in  a range $[-W,W]$
to the Kane-Mele Hamiltonian, and structural defects, such as constrictions across the ribbon width.  
We are interested in
coherent transport so we make use of the standard Landauer formalism as in Ref. \onlinecite{Fede06}: 
The disordered region contains $N=L_x\times L_y = 300$ lattice sites, where $L_y=10$ is the number of horizontal lines along the ribbon width and $L_x=30$ the number of atoms along each chain. 

\begin{figure}[h!]
\includegraphics[width=0.9\linewidth]{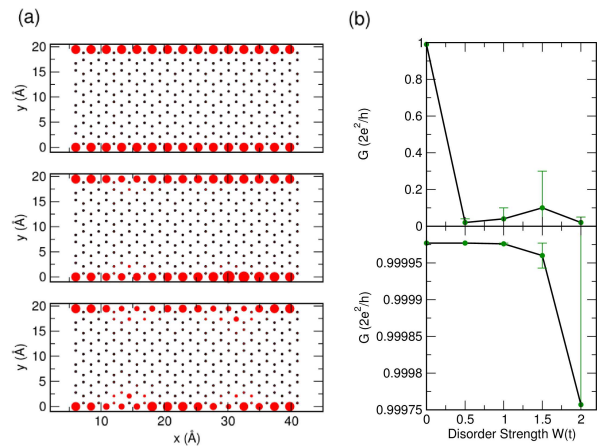}
  \caption{\label{charge-W}(\emph{Color online}) (a) Density of states at the Fermi energy 
projected on real space for a zigzag graphene nanoribbon with $t_{KM}=0.1t$  
and increasing values of the disorder strength (top) $W=0.0$, (middle) $W=1.0$ and (bottom) $W=2.0$.
(b) Conductance  
at the Fermi energy with (bottom) and without (top panel) SOC  for increasing disorder strength $W$. 
Error bars in conductance reflect statistical dispersion over disorder realizations.
}          
\end{figure}

In Figure \ref{charge-W}(a) we show the density of states at the Fermi level in real space for increasing 
disorder strength $W$ and for a SOC strength $t_\mathrm{KM}=0.1t$. 
All results with $W \ne 0$ have been averaged over 10 random realizations of disorder.
In the trivial insulator case for $t_\mathrm{KM}=0$, where a bulk gap arises due to lateral confinement, 
disorder induces intra-edge backscattering, localizing the edge states and 
strongly reducing the conductance at the Fermi level [see Fig.\ref{charge-W}(b)]. 
The effect of SOC on the conductance (close to the Fermi energy) is dramatic: even for large values of $W$, 
backscattering is almost entirely suppressed, as expected from their spin-filtered character.
 
\begin{figure}[h!]
\begin{center}
\includegraphics[width=0.9\linewidth]{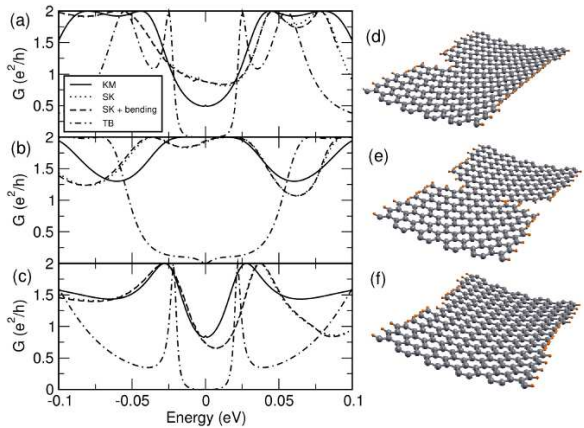}
\caption{ (\emph{Color online}) (\textbf{a-c}) Conductance for three different types of constrictions computed in different situations. Solid and dashed-dotted lines correspond to 
flat constrictions using the KM Hamiltonian with and without SOC, respectively. Dotted lines show the conductance for the same constrictions computed using SK approximation. Dashed lines are related to the same cases but with transverse bending as shown in panels (\textbf{d-f}), using the SK approximation.}
  
  \label{KMvsSK}
\end{center}
\end{figure}

In finite width systems, when spin-filtered states of different edges having the same spin polarization are able to interact one with each other, 
a finite inter-edge backscattering will be present and an exact quantization is no longer expected. 
We study this effect by introducing constrictions in the ribbon (see Fig. \ref{KMvsSK}, right panels). 
We model them by removing atoms in an area near the edge delimited by gaussian functions $ \beta e^{-\alpha x^2}$ 
($x$ runs along the edge) for random $\beta$ and $\alpha$.  Here we  take
$t_\mathrm{KM} = 0.01t$ or $\lambda = 4$ eV. In Fig. \ref{KMvsSK}  we show the conductance 
for three different constrictions as a function of energy. 
We can observe that, while in the absence of SOC (dashed-dotted lines) the conductance is negligible at the Fermi energy
because the edge states are able to interact with the states in the opposite edge and backscatter, for finite SOC the conductance increases 
appreciably (solid lines). These results have been obtained within the KM model.

L. Sheng {\em et al.} \cite{Haldane05} numerically showed that the QSHI phase survive in presence
of non conserving spin terms. Here we study this by introducing a
transversal bending over the constrictions previously analyzed. The curvature
introduces a non trivial angle for the spin polarization of the spin-filtered edge
states of graphene nanoribbons as mentioned in section \ref{terminations}. These
angles are different for each edge. This behavior raises the question whether
the relative phase between the spin orientation of the two channels could affect
the process of backscattering. 
In order to answer this question we need to use our 4-orbital SK
 Hamiltonian instead of the single orbital
KM model. First, we analyze the transport for flat constrictions and compare the results with the
KM ones. To be meaningful we need to saturate the $\sigma$ dangling bonds of the
carbon atoms with hydrogen atoms. We compute the two terminal conductance with
the ANT1D code, which is part of the ALACANT package\cite{ANT1D}. In Fig. \ref{KMvsSK}(a-c)
the dotted lines show the conductance for the three constrictions
considered. The difference between the SK and the KM conductance reveals the
approximation inherent to the KM model, but are essentially identical. While the
SOC term is exact in the SK formalism, the SOC introduced in the KM Hamiltonian
is a second order perturbative approach to the exact value which still preserves
electron-hole symmetry.\cite{,Dani2011}.

Finally we consider the same constrictions with a transverse bending, with a
radii of curvature of 4.1nm (see right panels in Fig. \ref{KMvsSK}). In this situation we expect that the
backscattering is less effective due to the fact that the spin polarization are
not parallel between the channels of opposite edges. The results are shown
by dotted and dashed lines, revealing that the process of
backscattering is the same for both situations, namely, for a flat ribbon where $s_z$ is
a good quantum number and for a curved ribbon where Rashba like terms are present. 

\section{The effect of Coulomb interactions}
\label{Hubbard}

In the previous section we have seen how edge states in the Kane-Mele model are robust with respect to non-magnetic weak disorder. In this section we address if,  as predicted\cite{Kane-Mele1,Kane-Mele2},  they are also also robust with respect to Coulomb interaction. For that matter,  we use  the Hubbard model within a mean field approximation (MFA), 
where the interaction term is approximated by $U\sum_i (n_{i,\uparrow} \langle n_{i_\downarrow} \rangle + n_{i,\downarrow} \langle  n_{i,\uparrow}) $.  More sophisticated approaches have also implemented recently\cite{DHL2011}.

\begin{figure}
[t]
\includegraphics[width=\linewidth,angle=0]{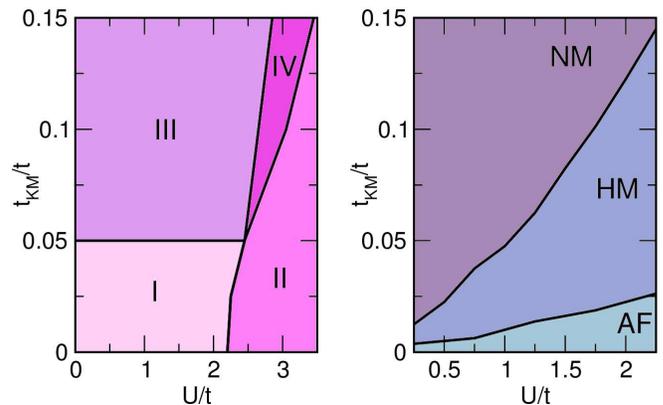}
\caption{ \label{figure8} 
(\emph{Color online}) On the left panel we show the phase diagram of an AC ribbon with $N_y=50$ in presence of Coulomb ($U/t$) and spin-orbit ($t_{KM}/t$) interactions. The labeled regions are related to the following phases: (I) Insulating or semiconducting phase, (II) antiferromagnetic insulating phase, (III) Quantum spin Hall insulating phase with gapless edge states and (IV) antiferromagnetic phase where metallic edge states are still preserved. On the right panel we show the phase diagram for the AF insulating (AF), the AF valley half-metal (HM) and the non-magnetic (NM) phases for the ZZ ribbon with $N_y=24$.}
\end{figure}

At half-filling,  the Hubbard model in the honeycomb lattice has magnetic order at sufficiently large $U$. In the MFA $U_c=2.2 t$, using Quantum Monte Carlo $U_c>5t$\cite{Sorella}. 
This magnetic ordering is always accompanied by a gap opening transforming graphene into a Mott  insulator.  On the contrary, the zigzag edge states have a very large density of states at the Fermi energy  so that,  at zero SOC,  undergo the ferromagnetic transition at arbitrary small $U$.  In the case of finite width ribbons the inter-edge correlations are anti-ferromagnetic and a band-gap opens in the otherwise conducting band structure\cite{Fujita96,Son06,Cohen06,JFR08}. 
 
Thus, under the separate influence of SOC and Coulomb repulsion, the preformed  zigzag edge states undergo a radical change in their electronic structure. In the first case they carry spin-currents and acquire a linear dispersion, in the second case they are ferromagnetic and, for finite width ribbons, open a gap.  Thus, it was not clear a priori what would be the combined effect of SOC and Coulomb repulsions.   This problem was addressed by two of us\cite{Soriano-JFR10}.  When the SOC is turned on, the effect on the insulting band structure of the magnetic phase is to decrease the band-gap in one valley and increase it in the other. This valley symmetry breaking goes together with a reduction of the edge magnetic moment.  As the SOC is increased, the gap closes in on of the valleys, the edges being still magnetic.  This phase is a valley half-metal.   As the SOC increases further, the magnetic moment goes to zero and the non-magnetic phase is recovered.   These findings are summarized in the phase diagram of  figure (\ref{figure8})b,  for a ZZ graphene ribbon containing $N_y=24$, that shows that, because of the SOC, a finite $U$ is needed to make the edges magnetic.

\begin{figure}
[t]
\includegraphics[width=\linewidth]{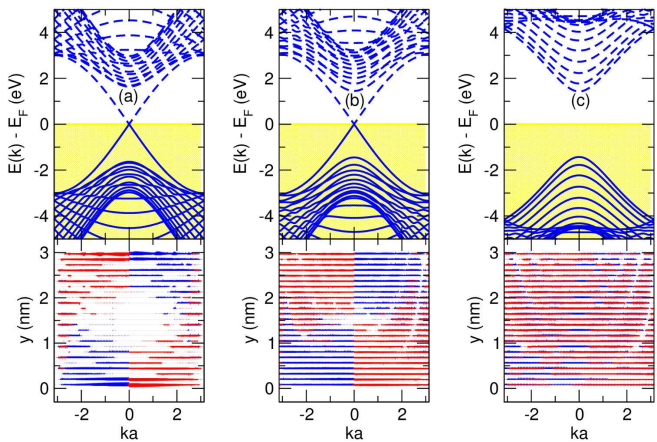} 
\caption{ \label{AC_SOC_bands} (\emph{Color online}) Band structure and projected spin density along the valence band of an AC ribbon with $N_y=50$ for three different states along the line $t_{KM}=0.1t$ in the phase diagram (a) ($U=1.0t$,$t_{KM}=0.1t$), (b) ($U=3.0t$,$t_{KM}=0.1t$) and (c) ($U=3.5t$,$t_{KM}=0.1t$)    
}
\end{figure}

We now consider the armchair termination, for which no preformed edges states exist, within the MFA for the Hubbard-Kane-Mele model.  This system  has also been addressed using bosonization techniques \cite{Sandler07}.
 In figure(\ref{AC_SOC_bands}) we show our results for the the band structure and edge-state wave function of an AC terminated ribbon, for 3 values of $U$, and $t_{KM}=0.1t$.  At the left, with $U=t$, the system is non-magnetic, and very similar to the $U=0$ case.  When the Coulomb interaction is increased above a critical value, the bulk atoms undergo an antiferromagnetic transition but the edge atoms stay non-magnetic and the spin-filtered edge states are preserved. Only when the interaction is increased above a $U_c$ significantly larger than the one that makes the bulk magnetic, the edge atoms go magnetic as well and the spin-filtered edge states dissappear.  Thus, for AC edges it is fair to stay that spin-filter edge states are robust with respect to Coulomb repulsion, since they survive even when the bulk goes antiferromagnetic. 
 
In figure (\ref{figure8})a we show the phase diagram of a semiconducting AC ribbon in presence of both Coulomb and spin-orbit interaction. For $t_{KM} < 0.05t$ and $U < U_c$ (region I) the ribbon remains semiconducting and for $U > U_c$ the gap increases and the ribbon becomes an antiferromagnet (region II). For $t_{KM} > 0.05t$ the ribbon transform into a QSHI holding spin filtered edge states as those shown in the case of the ZZ ribbons (region III) and, as the Coulomb interaction increases, we find a new phase (region IV) which is antiferromagnetic with a magnetic moment per atom in the edge of $|m| \sim 0.2$ but where the gap is still closed and the states remains spin-filtered.

\section{Summary}
\label{Summary}

We have studied the electronic properties of the spin-filtered edge states that appear in graphene because of spin-orbit coupling. Whereas they are a generic feature of the Quantum Spin Hall Insulator phase, their properties are very different depending on the existence of preformed edge states, {\it i.e.}, edge states that occur even with no SOC.  For instance, we reczag edges have 3  conductance channels,  and zigzag edges  are not robust with respect to Coulomb   interactions.   Whereas graphene is  not a good  material to study spin-filtered edge states experimentally, because of the low spin orbit coupling, it provides an ideal platform to study generic features of such an interesting object which can be found in other two dimensional topological insulators.

\begin{center}{\bf ACKNOWLEDGMENT }\end{center}

This work has been financially supported by MEC-Spain (MAT07-67845,FIS2010-21883) and CONSOLIDER CSD2007-0010).  We acknowledge fruitful conversations with L. Brey and T. Stauber and to the Centro de Computaci\'on Cient\'ifica of UAM for computational resources.

\end{document}